\documentclass[twocolumn]{aastex}
\usepackage{natbib}
\usepackage{graphicx}
\newcommand{\am}{\mbox{\arcmin}}
\newcommand{\as}{\mbox{\arcsec}}

\newcommand{\kms}{\mbox{\,km\,s$^{-1}$}}
\newcommand\cmv{\mbox{cm$^{-3}$}}
\newcommand\cmc{\mbox{cm$^{-2}$}}

\newcommand{\pp}{K cm$^{-3}$}

\newcommand{\lsun}{\mbox{L$_\odot$}}
\newcommand{\msun}{\mbox{M$_\odot$}}

\newcommand{\av}{\mbox{$A_V$}} 

\newcommand{\co}{$^{12}$CO}
\newcommand{\coo}{$^{13}$CO}

\newcommand{\hcop}{HCO$^+$}

\newcommand{\cplus}{C$^+$}

\def \CI{[C\,{\sc i}]}
\def \HI{H\,{\sc i}}
\def \CII{[C\,{\sc ii}]}

\def \HII{H\,{\sc ii}}

\def \beq{\begin{equation}}
\def \eeq{\end{equation}}
\begin{document}

\title {L1599B:  Cloud Envelope and \cplus\ Emission in a Region of Moderately Enhanced Radiation Field}
\author{Paul F. Goldsmith\altaffilmark{1}, Jorge L. Pineda\altaffilmark{1}, William D. Langer\altaffilmark{1}, Tie Liu\altaffilmark{2}, Miguel Requena--Torres\altaffilmark{3}, Oliver Ricken\altaffilmark{4}, and Denise Riquelme\altaffilmark{4}}

\altaffiltext{1}{Jet Propulsion Laboratory, California Institute of Technology, 4800 Oak Grove Drive, Pasadena CA, 
91109, USA; paul.f.goldsmith@jpl.nasa.gov}
\altaffiltext{2}{Korea Astronomy and Space Science Institute, 776 Daedeokdae-ro, Yuseong-gu, Daejeon 34055, Korea}
\altaffiltext{3}{Space Telescope Science Institute, 3700 San Martin Dr., Baltimore, 21218 MD, USA}
\altaffiltext{4}{Max--Planck--Institut f\"{u}r Radioastronomie, Auf dem H\"{u}gel 69, D-53121 Bonn, Germany}

\begin{abstract}

We study the effects of an asymmetric radiation field on the properties of a molecular cloud envelope.  We employ observations of carbon monoxide (\co\ and \coo), atomic carbon, ionized carbon, and atomic hydrogen to analyze the chemical and physical properties of the core and envelope of L1599B, a molecular cloud forming a portion of the ring at $\simeq$ 27 pc from the star $\lambda$ Ori.   The O III star provides an asymmetric radiation field that produces a moderate enhancement of the external radiation field.  Observations of the \CII\ fine structure line with the GREAT instrument on SOFIA indicate a significant enhanced emission on the side of the cloud facing the star, while the \CI, \co\ and \coo\ J = 1-0 and 2-1, and \co\ J = 3-2 data from the PMO and APEX telescopes suggest a relatively typical cloud interior.  The atomic, ionic, and molecular line centroid velocities track each other very closely, and indicate that the cloud may be undergoing differential radial motion.  The HI data from the Arecibo GALFA survey and the SOFIA/GREAT \CII\ data do not suggest any systematic motion of the halo gas, relative to the dense central portion of the cloud traced by \co\ and \coo.

\end{abstract}

\section{Introduction}

Interstellar clouds are bathed in an external radiation field produced primarily by stars, which affects the chemical and physical properties of the clouds' envelope.  For some aspects of cloud structure, such as the chemistry of molecular cloud cores, the external radiation field, particularly the ultraviolet (UV) component, may be ignored if there is sufficient dust opacity that the central portions of the cloud are shielded from the external radiation field.  For other aspects of cloud structure, such as the temperature and chemical composition of cloud boundaries, the external radiation field is critical.  

The regions of clouds in which the radiation field plays a major role are often referred to as Photon Dominated Regions, or PDRs \citep{Hollenbach99}.  While such regions studied are often intimately  associated with relatively massive young stars \citep[e.g., Orion;][]{Goicoechea15}, more modestly irradiated regions in which photons heat and dominate the ionization exist as well.  These include dark clouds such as Taurus, for which the radiation field may be even less than the ``standard'' interstellar radiation field \citep{Flagey09}.

The \CII\ fine structure line is an important probe of the interstellar medium, in part because it arises from ionized, atomic, and PDR regions.  Disentangling these multiple sources responsible for the 158 $\mu$m\ emission is critical for understanding how to interpret this spectral line and to evaluate its ability to serve as a tracer of star formation.  One important question is how the \CII\ intensity depends on the intensity of the external radiation field, which can vary enormously.  A study of a low-radiation field boundary region in Taurus \citep{Orr14} was able to detect only marginally the \CII\ line using the HIFI instrument \citep{degraauw10} on {\it Herschel} \citep{Pilbratt10}, with an intensity a factor $\simeq$\ 1000 less than that observed in the very intense UV radiation field characterizing the central portion Orion molecular cloud \citep{Goicoechea15}.  It  is thus valuable to study regions of intermediate external radiation field strength to test the ability of existing PDR models to predict emission from such environments, and to assess how the \CII\ emission varies as a function of radiation field.

The O8 star  $\Lambda$ Ori has produced a ring--like structure approximately 8\degr\ in angular diameter, which is seen optically as a fragmented, bright--rimmed obscuration around a largely empty interior around the star.  This region has long been observed in a variety of tracers starting with \HI\ \citep{Wade57}, H$\alpha$ \citep{Isobe73,Duerr82}, and radio continuum \citep{Reich78}.  The stellar content was studied by \citet{Murdin77}.   \citet{Maddalena87} and \citet{Lang98} observed CO throughout the region.  A more extensive compilation of references can be found in \citet{Maddalena87} and \citet{Lang2000}.

The ring is most visible in molecular line emission, but is also clearly delineated by the radio continuum emission that lies interior to it.  The situation of the atomic gas is less clear as different studies have yielded somewhat contradictory results about the extent of the \HI\ shell and its location.  There is little doubt that the interior of the molecular ring contains ionized gas  with $n_e$ = 2 \cmv\ \citep{Reich78}, and little or no atomic gas.  One explanation is that the ionized gas was produced by the O8 III star $\Lambda$ Ori together with its three B star companions \citep{Maddalena87}.  This study indicates that the luminosity of $\Lambda$ Ori alone is equal to 1.7$\times$10$^5$ \lsun.  However, this modeling effort recognized that the location and proper motion of $\Lambda$ Ori presented problems for this star having produced the ring through emission and mass loss starting $\simeq$ 2$\times$10$^6$ yr ago.  An alternative model is that a different star became a supernova which resulted in the ring that we now see \citep{Cunha96, Dolan01}.  

\begin{center}
\begin{figure}[h!]
\includegraphics[width = \columnwidth]{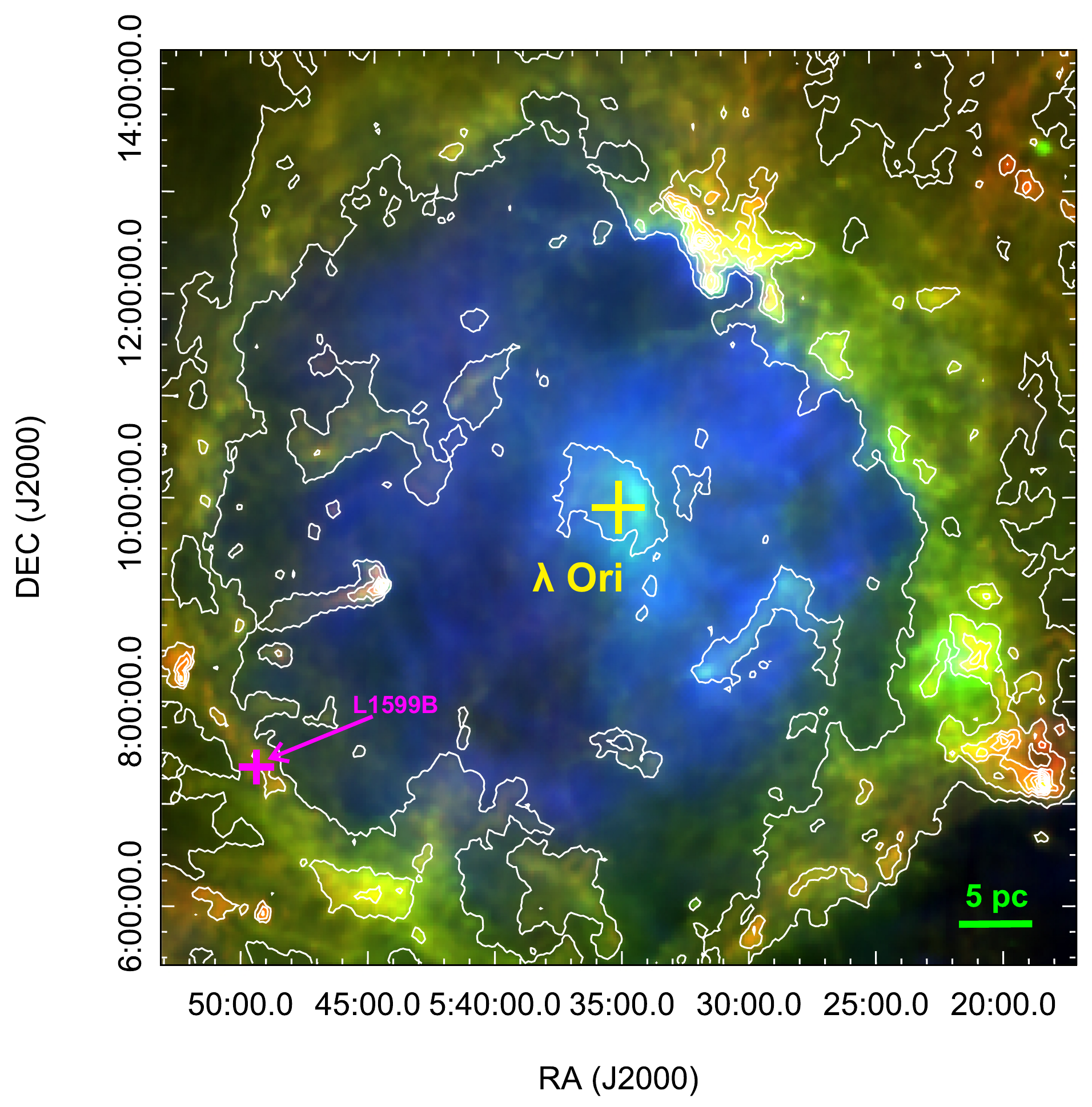}
\caption{\label{ringpix} 
Three--color composite image (Red: Planck 857 GHz; Green: IRAS 100 $\mu$m; Blue: H$\alpha$) of the $\Lambda$ Orionis region. 
The white contours represent the flux density of Planck 857 GHz continuum emission. 
The contours are from 10\% to 90\% in steps of 10\% of the peak value, which is 133.1 MJy sr$^{-1}$.
This figure has been adapted from \citet{Liu16}.
}
\end{figure}
\end{center}

The most massive cloud in the $\Lambda$ Ori ring is B30, while the bright--rimmed cloud B35 projects from the ring towards the central star.  The L1599 region is number 11 in Fig. 3 of \citet{Maddalena87} and number 8 in Fig. 2 of \citet{Lang2000}.  Figure \ref{ringpix} gives an overview of the $\Lambda$ Ori ring, with the magenta cross denoting L1599B.  The L1599B cloud was one of those observed in CO and \HI\ by \citet{Wannier83}, who were investigating an apparent \HI\ halo surrounding this obect.  Irrespective of the details of the origin of the $\Lambda$ Ori ring, L1599B and other clouds on the ring's periphery are bordered on the one side by the \HII\ region, which is kept ionized by the radiation from the O-- and B--stars.  Consequently, the cloud is subjected to a highly asymmetric radiation field, which makes it an ideal environment for studying the radiation field dependence of the interface between molecular and atomic/ionized gas regions.  

In this paper we report an observational study of the L1599B cloud. \CII\ emission was observed with the GREAT\footnote{GREAT is a development by the MPI f\"ur Radioastronomie and the KOSMA/ Universit\"at zu K\"oln, in cooperation with the MPI f\"ur Sonnensystemforschung and the DLR Institut f\"ur Planetenforschung.}  instrument \citep{Heyminck12} on board SOFIA\footnote{This work is based in part on observations made with the NASA/DLR Stratospheric Observatory for Infrared Astronomy (SOFIA). SOFIA is jointly operated by the Universities Space Research Association, Inc. (USRA), under NASA contract NAS2-97001, and the Deutsches SOFIA Institut (DSI) under DLR contract 50 OK 0901 to the University of Stuttgart.}  \citep{Young12}.  
The \co\ J = 3--2 transition, the $^3$P$_1$-$^3$P$_0$ transition of \CI, and the J = 2--1 transition of \co\ and \coo\ were observed using the APEX telescope\footnote{This publication is based in part on data acquired with the Atacama Pathfinder Experiment (APEX). APEX is a collaboration between the Max-Planck-Institut f\"ur Radioastronomie, the European Southern Observatory, and the Onsala Space Observatory.} \citep{Guesten06}.  
The \co\ and \coo\ 1-0 lines were observed with the 13.7 m Purple Mountain Observatory (PMO) telescope.  
In \S\ \ref{obs} we describe the SOFIA, APEX, and PMO observations, and 
in \S\ \ref{data}, we present the observational data and fitted spectral line parameters.
In \S\ \ref{env} we discuss the environment of L1599B and derive the column densities of the various tracers.
In \S\ \ref{discussion} we discuss the cloud kinematics and evaluate the structure of the cloud boundaries, including comparing a two--sided slab PDR model with the data, and 
in \S\ \ref{summary} we summarize the results of our study.

\section{Observations}
\label{obs}

\begin{figure*}
\centering
\includegraphics[width = 0.9\textwidth]{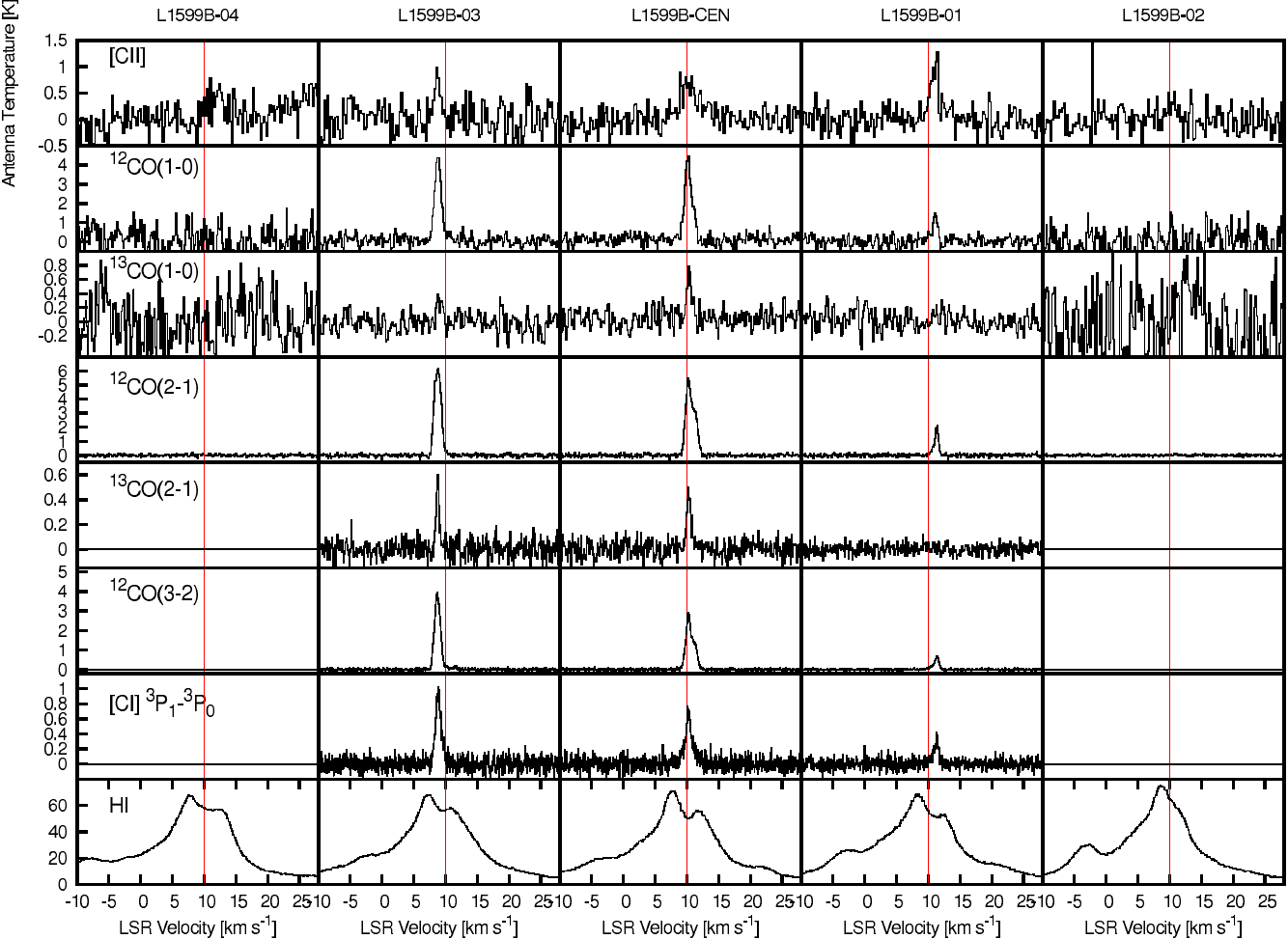}
\caption{\label{spectra} Spectra of ionic, atomic, and molecular species observed towards the five positions in the strip through L1599B.  The species and transition are identified in the left--most column (corresponding to position 4).  The spectra are presented in units of antenna temperature corrected for atmospheric absorption.  The vertical red line denotes a velocity of 10 \kms, highlighting the systematic velocity shifts that occur in all species as one moves across the cloud.  Positions 01 and 02 are in the direction towards the star $\Lambda$ Ori.}
\end{figure*}

We observed five positions listed in Table \ref{table_pos}, comprising a strip (more or less oriented on a line pointing towards $\lambda$ Ori) across the narrow axis of L1599B.
The ``Center'' (CEN) position is approximately in the middle of the strip, and was chosen based on the data from \citet{Wannier83}. Figure \ref{spectra} shows the spectra obtained at the five positions across the strip.
The frequencies of the observed transitions are given in Table \ref{table_freqs}.

\begin{table}[!t]
\caption{Positions (J2000) Observed in L1599B}
\begin{center}
\begin{tabular}{ccc}
\hline \hline
Designation&	Right Ascension&	Declination\\
\hline
Center (CEN)&05:49:27.40	& +07:22:07.0\\
O1		&05:49:03.80 	& +07:29:46.0\\
O2		&05:48:19.60 	& +07:37:42.7\\
O3		&05:49:48.57 	& +07:16:25.0\\
O4		&05:50:34.51	& +07:04:24.7\\
\hline
\end{tabular}
\end{center}
\label{table_pos}
\end{table}

\begin{table}[!t]
\caption{Frequencies of Transitions Observed}
\begin{center}
\begin{tabular}{ccr}
\hline \hline
Species&	Transition &	Frequency (GHz)\\
\hline
\hcop &1 -- 0    &   89.1885\\
\coo   &1 -- 0      & 110.2014\\
\co &1 -- 0      & 115.2712\\
\coo   &2 -- 1     &220.3987\\
\co &2 -- 1	&230.5380\\
\co   &3 -- 2	&345.7960\\
\CI	&$^3$P$_1$ -- $^3$P$_0$ &492.1607\\
\CII &$^2$P$_{3/2}$ -- $^2$P$_{1/2}$ &1900.5369\\
\hline
\end{tabular}
\end{center}
\label{table_freqs}
\end{table}

\subsection{\hcop}
\label{HCO+}
We observed the five positions given in Table \ref{table_pos} with the 21 m Yonsei telescope \citep{Kim11} of the Korean VLBI Network (KVN) on 7 September 2015.  The FHWM beam size and beam efficiency were 32\arcsec\ and 0.41, respectively.  Smoothing the data to 0.42 \kms\ yielded an rms antenna temperature uncertainty of 0.03 K.  A signal above the noise level was seen only at the CEN position, where the peak line intensity was 0.11 K, the line width 2.3 \kms, and the central velocity 10.1 \kms.  Despite the marginal ($\simeq$\ 5$\sigma$\ in integrated intensity) detection, the line parameters agree well with those of the other species detected at this position, so we do feel this represents a detection of \hcop.  We discuss the significance of this observation in terms of possible high density component of the cloud in \S\ref{kinematics}.

\subsection{CO}

\begin{center}
\begin{figure}
\includegraphics[width = 0.8\columnwidth,angle = -90]{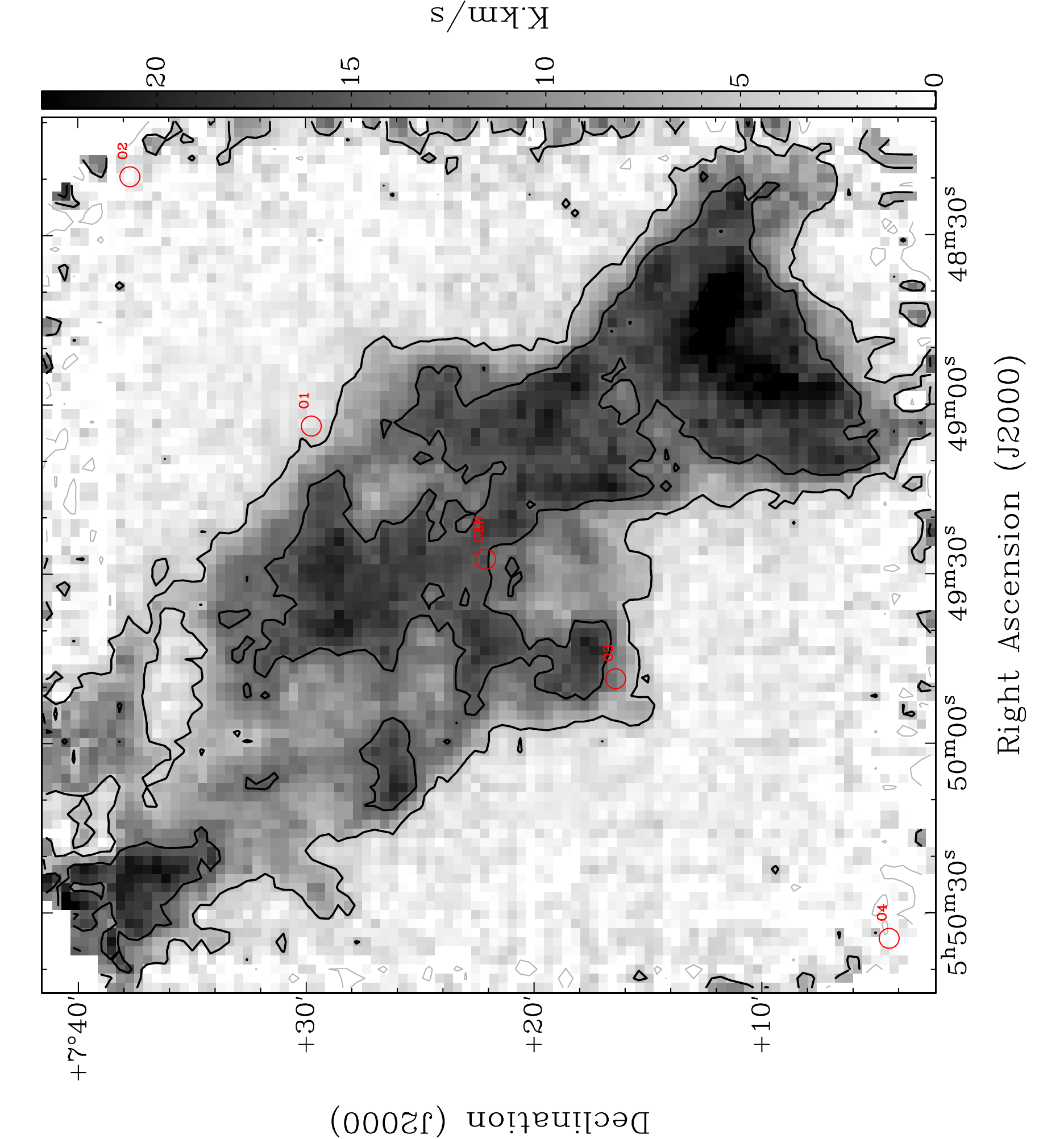}
\caption{\label{12CO_PMO}Image of \co\ J = 1 -- 0 main beam temperature in the L1599B cloud integrated over the velocity range 0 to 20 \kms.  The contours are at 30\% to 90\% of the maximum integrated temperature in steps of 20\%.  The beam size is $\simeq$ 50\as\ FWHM.  The positions at which we make a detailed comparison with other species are indicated by the designations given in Table \ref{table_pos}.}
\end{figure}
\end{center}

\begin{center}
\begin{figure}
\includegraphics[width = 0.8\columnwidth,angle = -90]{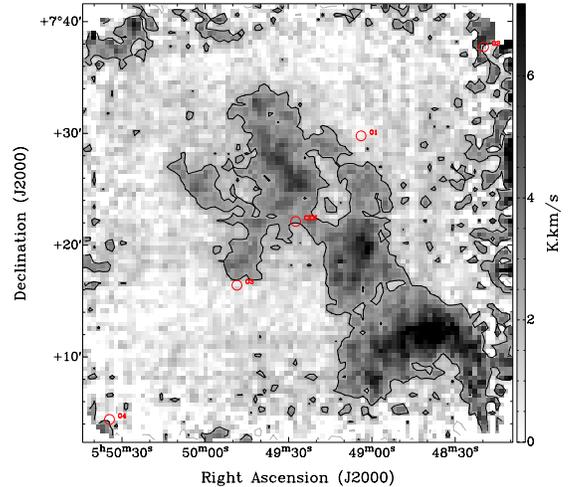}
\caption{\label{13CO_PMO}Image of \coo\ J = 1 -- 0  in the L1599B cloud.  Notes as for Fig. \ref{12CO_PMO}. }
\end{figure}
\end{center}

The $^{12}$CO and \coo\ J = 1--0 lines were observed with the 9--beam receiver \citep{Shan12} on the Purple Mountain Observatory 13.7 m telescope on 15 October 2015.  The data were converted to a main beam temperature scale by using main beam efficiencies of 0.52 for 115 GHz and 0.56 for 110 GHz (T. Liu, private communication).  The FWHM beam size was $\simeq$ 50\as.  The spectra shown in Figure \ref{spectra} were part of small maps of the region in each isotopologue.  Figure \ref{12CO_PMO} displays the \co\ image and Figure \ref{13CO_PMO} the \coo\ image.  The locations of the 5 positions studied in detail are also indicated in these figures. 

The higher--J carbon monoxide observations were carried out using 
the 12 m  Atacama Pathfinder EXperiment (APEX) telescope \citep{Guesten06} on several nights in September 2014 under good weather conditions.  
The J = 2--1 rotational transition of \co\ and \coo\ were observed using the APEX-1 (SHFI) receiver \citep{Vassilev08} and the extended bandwidth Fast Fourier Transform Spectrometer (XFFTS) backend \citep{Klein12}.  
The J = 3--2 rotational transition of \co\ was observed with the FLASH+ receiver \citep{Klein14} also on APEX and the XFFTS backend.
Only the center and the two closest offset positions were observed in all of the lines after getting nondetections at the outer positions. 
Position switching observations with a clean reference position  ($\alpha$(J2000) = 05:52:30, $\delta$(J2000) = 06:39:00) were used to achieve the observed signal to noise ratio. For both sets of observations, the focus was checked at the beginning of each observation; pointing was checked every hour and found to be better than 2$''$. 
The data from APEX were calibrated in units of T$^*_a$ using the standard APEX calibration tools, which provide a error of $\simeq$10\%.
A beam efficiency of 0.76 was used for analysis of the J = 2--1 \co\ and \coo\ lines observed with SHFI and 0.69 for the J = 3--2 \co\ lines observed with FLASH+.  The FWHM beam sizes were 27\as\ to 28\as\ for the J = 2--1 observations and 17.5\as\ for the J = 3--2 observations. 

\subsection{\CI}

The \CI\ $^3$P$_1$ -- $^3$P$_0$ observations were carried out using the APEX telescope  in parallel with the CO J = 3-2 observations discussed above, using the FLASH+ receiver. The FWHM beam size was 12.4\as.  A beam efficiency of 0.57 was used for analysis of the \CI\ data.

\subsection{\CII}
Observations of the 1900.537 GHz $^2$P$_{3/2}$ -- $^2$P$_{1/2}$ \CII\ fine structure line were carried out as part of SOFIA Cycle 1 project  01$_{-}$0040US on 2 November 2013, using the GREAT \citep{Heyminck12} instrument on the SOFIA airborne observatory \citep{Young12}.     The observations were carried out at 43000 ft (13.1 km) altitude under excellent observing conditions and with an average system  temperature of 2970 K (SSB). 
We tuned the receiver to the 1900.537 GHz \CII\ line frequency in the upper sideband (USB). 
We employed a digital FFT spectrometer \citep{Klein12} that analyzed a bandwidth of 1.5 GHz with a spectral resolution of 0.212 MHz (0.034 \kms).  

The L1599B positions were observed in Position Switching mode, using two different reference positions free of emission depending on the offset; (05:51:37.9, 07:28:16.0)  for the Cen, 03, and 04 positions, and (05:47:30, 07:37:59) for the 01 and 02 positions.
The nominal focus position was updated regularly based on measurements of the telescope structure. The pointing was established with the optical guide cameras to an accuracy of better than a few arcsec.
The forward and main beam efficiencies were 0.97 and 0.67, respectively, and the FWHM beam size 15\as.
The data were calibrated with the KOSMA/GREAT calibration procedure \citep{Guan12}, removing residual telluric lines. The calibrated 
data were reduced with the CLASS\footnote{http://www.iram.fr/IRAMFR/GILDAS} software removing a first order polynomial baseline from an rms--weighted average of the data.

\subsection{\HI}

The \HI\ data were obtained from the GALFA survey \citep{Peek11}, carried out with the Arecibo 305 m telescope.  The angular resolution is 4\am\ and the velocity resolution 0.2 \kms.

\section {Data}
\label{data}

Table \ref{datatable} gives the basic parameters of the spectral lines observed at the five positions in the strip through L1599B. The line parameters are obtained from Gaussian fits of one or two components as appropriate for the spectrum in question.  The \HI\ is not included as Gaussian fitting is not helpful due to the complex line profiles with multiple emission components as well as possible self--absorption (discussed further in \S\ \ref{HI}).
%
%
\begin{table*}
\tiny
\caption{Observational Data\tablenotemark{a}  on L1599B}
\label{datatable}
\begin{tabular}{ccccccc}
\hline \hline
Spectral Line&Position&04&03&CEN&01&02\\
\hline
\CII\ $^2$P$_{3/2}$--$^2$P$_{1/2}$ &$T_{A~\rm max}$              &0.42$\pm$0.07  &0.56$\pm$0.26&0.37$\pm$0.07&0.63$\pm$0.11&0.1$\pm$0.1\\
&$T_{A~\rm int}$                 &0.50$\pm$0.15  &0.70$\pm$0.15&1.25$\pm$0.13&1.17$\pm$0.13&0.1$\pm$0.1\\ 
&$V_{\rm lsr}$                     &12.4$\pm$0.2   &8.7$\pm$0.12   &10.3$\pm$0.17&11.0$\pm$0.1   &\\
&$\delta V_{\rm FWHM}$     &1.1$\pm$0.4    &1.2$\pm$0.3     &3.1$\pm$0.4    &1.8$\pm$0.3    &\\
&$T_{A~\rm max}$		     &0.25$\pm$0.11 &--&--&--&--\\
&$T_{A~\rm int}$                 &0.38$\pm$0.16  &--&--&--&--\\
&$V_{\rm lsr}$                      &10.6$\pm$0.3   &--&--&--&--\\
&$\delta V_{\rm FWHM}$    &1.4$\pm$0.6      &--&--&--&--\\
\hline
$^{12}$CO J = 1--0     &$T_{A~\rm max}$         &--& 4.32$\pm$0.34   & 1.15$\pm$0.24   &  1.50$\pm$0.23   &  -- \\
&$T_{A~\rm int}$            &--& 5.60$\pm$0.18   & 0.57$\pm$0.17   &  1.18$\pm$0.13   & -- \\  
&$V_{\rm lsr}$                &--& 8.79$\pm$0.02   &11.27$\pm$0.05 & 11.14$\pm$0.04  &  -- \\
&$\delta V_{\rm FWHM}$&--& 1.22$\pm$0.05 & 0.52$\pm$0.07  &  0.74$\pm$0.08   &  -- \\
&$T_{A~\rm max}$         &--&--                          &  4.23$\pm$0.24  &  0.47$\pm$0.23   &  -- \\
&$T_{A~\rm int}$           &--&--                           &  5.64$\pm$0.24  &  0.28$\pm$0.11   & -- \\ 
&$V_{\rm lsr}$               &--&--                           &10.16$\pm$0.02 &10.08$\pm$0.11   & --  \\
&$\delta V_{\rm FWHM}$&--&--                         &  1.25$\pm$0.07 &  0.55$\pm$0.28   & --  \\
\hline
$^{13}$CO J = 1--0        & $T_{A~\rm max}$          &--& 0.35$\pm$0.12 & 0.68$\pm$0.13     &--                           &--\\
&$T_{A~\rm int}$             &--& 0.40$\pm$0.07 & 0.66$\pm$0.08     &--                           &--\\ 
&$V_{\rm lsr}$                 &--& 9.04$\pm$0.10 &10.36$\pm$0.05    &--                           &--\\
&$\delta V_{\rm FWHM}$&--& 1.07$\pm$0.19 & 0.91$\pm$0.14     &--                           &--\\
\hline
$^{12}$CO J = 2--1&$T_{A~\rm max}$              &--&5.86$\pm$0.09&3.01  $\pm$0.09&1.76   $\pm$0.09&0.09$\pm$0.03\\
&$T_{A~\rm int}$                 &--&6.49$\pm$0.38&3.39  $\pm$0.15&1.02   $\pm$0.13&0.074$\pm$0.029\\
&$V_{\rm lsr}$                     &--&8.93$\pm$0.03&11.29$\pm$0.02&11.39  $\pm$0.01&11.01$\pm$0.13\\
&$\delta V_{\rm FWHM}$    &--&0.68$\pm$0.04&1.06  $\pm$0.03&0.54   $\pm$0.02&0.77$\pm$0.45\\
&$T_{A~\rm max}$              &    &2.57$\pm$0.09&5.19$\pm$0.09&0.56   $\pm$0.09&--\\
&$T_{A~\rm int}$                 &     &1.87$\pm$0.36&5.52$\pm$0.15&0.61  $\pm$0.15&--\\ 
&$V_{\rm lsr}$                     &     &8.27$\pm$0.03&10.20$\pm$0.01&10.92$\pm$0.13&--\\
&$\delta V_{\rm FWHM}$    &   &0.68$\pm$0.04&1.00   $\pm$0.02&1.02  $\pm$0.14&--\\
\hline
$^{13}$CO J = 2--1&$T_{A~\rm max}$            &-- --&0.56$\pm$0.07&0.43$\pm$0.64   &--&-- --\\
&$T_{A~\rm int}$               &-- --&0.34$\pm$0.03&0.40$\pm$0.03   &--&-- --\\ 
&$V_{\rm lsr}$                   &-- --&8.76$\pm$0.02&10.31$\pm$0.03 &--&-- --\\
&$\delta V_{\rm FWHM}$  &-- --&0.57$\pm$0.05&0.86$\pm$0.08   &--&-- --\\
\hline
$^{12}$CO J = 3--2&$T_{A~\rm max}$  	          &-- --&3.88$\pm$0.05         &1.29$\pm$0.04    &0.45$\pm$0.03  & -- --\\
&$T_{A~\rm int}$                &-- --& 4.46$\pm$0.01        &1.28$\pm$0.02     &0.27$\pm$0.01  &-- --\\
&$V_{\rm lsr}$                    &-- --& 8.73$\pm$0.01        &11.25$\pm$0.008  &11.41$\pm$0.01&-- --\\
&$\delta V_{\rm FWHM}$   &-- --&1.08$\pm$0.04         &0.94$\pm$0.01      &0.56$\pm$0.02 &-- --\\
&$T_{A~\rm max}$             &-- --&--                              &2.73$\pm$0.04      &0.27$\pm$0.03       &-- --\\
&$T_{A~\rm int}$                &-- --&--                              &2.73$\pm$0.02      &0.42$\pm$0.01&-- --\\ 
&$V_{\rm lsr}$                    &-- --&--                              &10.22$\pm$0.01    &11.06$\pm$0.01&-- --\\
&$\delta V_{\rm FWHM}$   &-- --&--                              &0.94$\pm$0.01      &1.47$\pm$0.05&-- --\\
\hline
\CI\ $^3$P$_1$--$^3$P$_0$ &$T_{A~\rm max}$          &-- --&0.90$\pm$0.08&0.60$\pm$0.09&0.28$\pm$0.06   &-- --\\
&$T_{A~\rm int}$             &-- --&1.08$\pm$0.03&0.87$\pm$0.03&0.29$\pm$0.02   &-- --\\ 
&$V_{\rm lsr}$                 &-- --&8.89$\pm$0.01&10.28$\pm$0.02&11.27$\pm$0.03&-- --\\
&$\delta V_{\rm FWHM}$&-- --&1.13$\pm$0.04&1.38$\pm$0.07&0.97$\pm$0.09    &-- --\\
\hline 
\end{tabular}

\tablenotetext{a} {Peak intensities are antenna temperature (K), integrated intensities are antenna temperature integrated over velocity (K \kms), and  velocities and line widths are in \kms.  For positions where two Gaussian components were indicated, the two sets of fitted parameters are given sequentially.  Positions where observations were not carried out are indicated by double dashes (-- --).  Positions where no detection was made (or only a single component could be fitted) are indicated by a single dash (--).  The velocity and line width of the \CII\ line at position 02 could not be reliably determined.  The identification of two components at position 04 is marginal but is a better fit than a single wide component.}
\end{table*}


\section{Cloud Environment and Column Densities}
\label{env} 

\subsection{Environment}
We adopt a distance of 425 pc for L1599B and $\Lambda$ Ori.  This distance is slightly less than the value of 445 pc suggested by \citet{Lombardi11}, slightly greater than the 400 pc used by \citet{Maddalena87}, but close to the distance to the clouds in the $\Lambda$ Ori ring of 420 pc derived by \citet{Schlafly14}.  

The radiation flux at L1599B is enhanced by $\Lambda$ Ori and the lower--mass stars in the Collinder 69 stellar cluster.  The mass of $\Lambda$ Ori is 26.8 \msun\ \citep{Dolan01}.  From Table 1 of \citet{Parravano03}, we find that the star's output in the FUV-H$_2$ range (912-2070 \AA) is 2.2$\times$10$^{38}$ erg s$^{-1}$.  For a distance from $\Lambda$ Ori to L1599B of 27 pc, we find a flux in the FUV-H$_2$ range of about 1.6 times that of the Habing field.  Given the contributions of the other moderately luminous stars in the cluster \citep{Murdin77}, we adopt an enhancement factor of 5 for modeling the physics and chemistry of the surface layers of L1599B facing $\Lambda$ Ori.  This may be an overestimate due to dust extinction between the star and the cloud.  The oppositely oriented surface of L1599B is subject to the standard interstellar radiation field.  

\subsection{Column Densities}
\label{coldens}
\subsubsection{Carbon Monoxide}
\label{co}

We have measurements of the three lowest transitions of \co\ at the CEN and 03 positions.  From each line, assuming that the transition is optically thick, we can derive the excitation temperature. The stronger component at the CEN position is at 10.2 \kms\ (in all three transitions), and to obtain the excitation temperature we employ the  expression for the antenna temperature produced by an optically thick line \citep[e.g.][]{Penzias75, Garden91}, expressed as
\beq
T_{\rm ex} = \frac{T^*}{ln(1 + \frac{T^*}{T_{\rm mb} + B})} ,
\eeq
where $T^*$ = $hf/k$ with $f$ the frequency of the transition. $T_{\rm mb}$ is the source minus reference main beam peak temperature and $B$ = $T^*/(exp(T^*/T_{\rm bg}) -1)$, where $T_{\rm bg}$ = 2.73 K is the background temperature.  
From these data we obtain $T_{ex}$ = 11.4, 11.8, and 9.7 K for the 1--0, 2--1, and 3--2 transition, respectively, for the CEN position. The excitation temperatures for the dominant 8.7--8.9 \kms\ component at the 03 position for the three transitions is 10.9, 12.8, and 12.4 K, respectively.  The three temperatures for each position are in quite close agreement, and suggest that we can adopt mean values of 11 K for CEN and 12 K for 03. With the additional assumption that the three lines are all thermalized, this leads to kinetic temperatures, $T_{\rm k}$, between 11 and 12 K in the region where the \co\ becomes optically thick.  As will be seen below, this solution is critically dependent on there being sufficient radiative trapping for this isotopologue. 

We have detected the two lowest transitions of \coo\ at the CEN and 03 positions.  For this species, the ratio of the two lowest transitions is a good tracer of the volume density, especially when the emission is optically thin, as we find is the case for L1599B.  From Table \ref{datatable}, we find that the ratio of the integrated main beam temperatures, $R(2/1)$, is 0.61 for the CEN position and 0.85 for the 03 position, with errors of approximately 20\%.  The ratio is directly expressible in terms of the ratio of the upper level column densities, which in turn can be expressed in terms of the excitation temperature of the J = 2-1 transition
\beq
R(2/1) = \frac{\int T_{\rm mb} (J = 2-1) dv}{\int T_{\rm mb}(J = 1-0) dv} = 4{\rm e}^{-10.6/T_{\rm ex}(J = 2-1)},
\eeq
where 10.6 K is the energy difference between the J = 2 and J = 1 levels expressed as a temperature.  We see that to obtain the observed ratios, excitation temperatures are very low; $\simeq$ 5.6 K for CEN and $\simeq$ 6.8 K for 03.  Thus a LTE population distribution cannot be assumed for calculating the \coo\ emission.

For the moderate radiation fields present here, the hydrogen will be almost entirely molecular (see e.g. discussion in \S\ \ref{edges}). Using the RADEX code \citep{vandertak07}, we find that solutions for $R(2/1)$, given the above kinetic temperatures, are restricted to the range 1000 $\leq$ $n({\rm H}_2$) $\leq$ 3000 \cmv.  
We find the closest agreement between data and model with $T_{\rm k}$ = 12 K, $n({\rm H}_2)$ = 1200 \cmv, and $N(^{13}$CO) = 1.3$\times$10$^{15}$ \cmc\ for the CEN position, while for the 03 position,  $T_{\rm k}$ = 11 K, $n({\rm H}_2)$ = 2500 \cmv, and $N(^{13}$CO) = 7$\times$10$^{14}$ \cmc.

With these \coo\ column densities and assuming a \co/\coo\ abundance ratio of 65 \citep{Langer93,Liszt07}, we find \co\ column densities between 5 and 9 $\times$10$^{16}$ \cmc.  Taking a representative value of 6$\times$10$^{16}$ \cmc\ and a line width of 1 \kms, with $n({\rm H}_2$) = 2000 \cmc\ and $T_{\rm k}$ = 12 K gives optical depths of 8, 17, and 11 for the three lowest \co\ transitions.  These are sufficiently large to give excitation temperatures of 11.0, 10.4, and 8.8 K.  The assumption that the \co\ yields the kinetic temperature is thus confirmed for the two lowest transitions, but is somewhat marginal for the J = 3-2 transition.  This difference may suggest that that the single density model is not correct in detail, but that using the kinetic temperatures derived above for analysis of the carbon monoxide emission should be adequate for the present discussion.

\subsubsection{Atomic Carbon}
\label{CI}

\begin{center}
\begin{figure}[h!]
\includegraphics[width = 1.0\columnwidth,angle = 0.0]{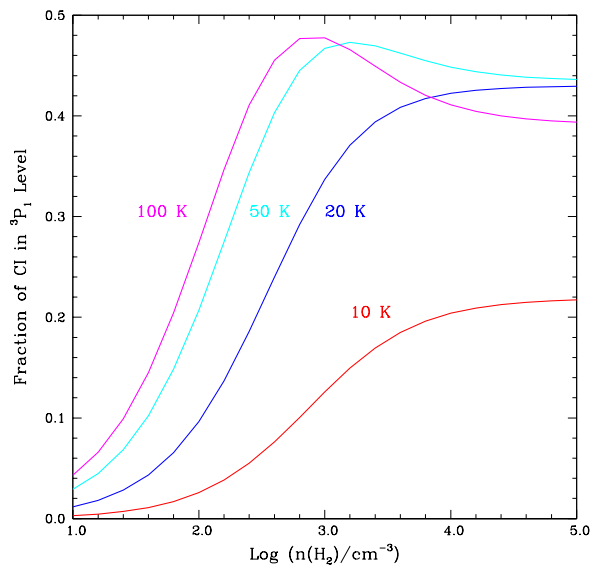}
\caption{\label{CI_fracpop} Fractional population of the $^3$P$_1$ fine structure level of atomic carbon for selected kinetic temperature (denoted by differently--colored lines) as a function of molecular hydrogen density.  The optical depths of both \CI\ transitions are assumed to be small.}
\end{figure}
\end{center}

We have detected the \CI\ 492 GHz transition at the 01, CEN, and 03 positions.  With the reasonable (and finally self--consistent) assumption of optical thinness, the upper level ($^3$P$_1$) column density is given by $N_u$ = [$8\pi\times10^5 k \nu^2/A_{ul}hc^3] \int T_{\rm mb}dv$ (K \kms) = 6.0$\times$10$^{15}\int T_{\rm mb}dv$ (K \kms).  The total column densities depend on the excitation conditions; the three level problem can be solved analytically yielding the fractional population of each level \citep[e.g.][]{Goldsmith15}.  The results for the $^3$P$_1$ level are shown in Figure \ref{CI_fracpop}.  In a cloud with modest total column density  such as L1599B, there will be a significant atomic carbon abundance throughout the cloud, and while it may be depressed just at the center due to conversion to carbon monoxide, the abundance of C will not be enhanced in the warm, outer layers, where the carbon is primarily \cplus.  

In considering the results shown in Figure \ref{CI_fracpop}, we see that the fractional population of $^3$P$_1$ is between 0.15 and 0.45.  If we assume a characteristic value of 0.3, which is the value appropriate for 20 K and $n({\rm H}_2$) = 1000 \cmv, we will be making an error of less than a factor of 2.  With the assumption that the H$_2$ density in the region producing the \CI\ emission is the same as that determined from \coo, we find atomic carbon column densities of 1.0, 2.9, and 3.6 $\times$10$^{16}$ \cmc\ for the 01, CEN, and 03 positions, respectively.  

\subsubsection{Ionized Carbon}
\label{cplus}

The $^2$P$_{3/2}$-$^2$P$_{1/2}$ fine structure line of \cplus\ (\CII) was detected at all 5 positions observed, albeit with limited signal to noise ratio, particularly at the 02 position.  The line profiles at the three inner positions are surprisingly similar to those of both the molecular and atomic species in terms of line width and central velocity.  This similarity indicates that the conditions and extent of the emitting region are not vastly different from those of the other species observed.  

In order to determine the excitation and column density of \cplus, we  assume that the thermal pressure in the region responsible for the \CII\ emission is the geometric mean of the ISM pressure and the pressure of the internal molecular zone \citep{Wolfire03,Pineda13}.  For the ISM pressure, we adopt the mean value of the range found by \citet{goldsmith13}, 5700 \pp, determined for translucent clouds with low extinctions, which are assumed to be in pressure equilibrium with the surrounding ISM.  For the molecular zone pressure, we use the average of the values found at the CEN ($p/k$  = 18,000 \pp) and 03 ($p/k =$27,500 \pp) positions discussed above, yielding $\overline{p/k}$ = 20,950 \pp.  Taking the geometric mean yields a thermal pressure in the \cplus\ layer of 10,900 \pp.  Based on expectations of cloud--edge thermal structure, we adopt conditions n(H$_2$) = 100 \cmv, and T$^{\rm k}$ = 100 K for analyzing the \CII\ emission.  

Inverting  equation 26 of \citet{Goldsmith12} yields
\beq
N({\rm C}^+) = 2.9\times10^{15}[1 + 0.5e^{91.2/T^{\rm k}}(1 + \frac{2.4\times 10^{-6}}{nR_{\rm ul}})]\int T_{\rm mb} dv ~,
\eeq
where the column density is in \cmc, the integrated temperature in K\kms, $n$ is the volume density of colliding particles in \cmv, and $R_{\rm ul}$ is the collisional deexcitation rate in cm${^3}$s$^{-1}$.  Throughout almost all the \cplus\ region, hydrogen will be molecular, so we use the $R_{\rm ul}$ rate coefficient from \citet{Wiesenfeld14}, assuming a LTE H$_2$ ortho to para ratio.  For 100 K kinetic temperature $R_{\rm ul}$ = 5.1$\times$10$^{-10}$ cm$^3$ s$^{-1}$, and we obtain $N({\rm C}^+) = 1.8\times 10^{17} \int T_{\rm mb} dv$.  The C$^+$ column densities are 2.4, 1.9, 3.4, 3.2, and 0.3 $\times$10$^{17}$ cm$^{-2}$ at positions 04, 03, CEN, 01, and 02, respectively.

\subsubsection{Atomic Hydrogen}
\label{HI}

Analysis of the \HI\ data is problematic due to the difficulty of associating atomic gas specifically with the L1599B region, given that the \HI\ emission is widespread throughout the $\Lambda$ Ori ring as indicated by early observations of \citet{Wade57}.  The extended \HI\ is confirmed by the GALFA observations, from which the spectra in Figure \ref{spectra} were extracted.  There are multiple velocity components present, as seen in Figure \ref{spectra}, but also one at -20 \kms.  The velocities of peak emission do shift slightly with position, presumably reflecting the complex motions of gas in the expanding ring \citep[e.g.][]{Maddalena87}.  

In the region of L1599B studied in detail, the \HI\ profiles are unusual in that the spectra at the three central positions have two maxima, which could reflect the presence of two velocity components, or could be the result of \HI\ self--absorption.  If we adopt the latter interpretation, the line widths of the absorption features are between 1 and 2 \kms, similar to that of the molecular emission lines, and the velocities agree very well.  Thus, this can be considered to be an example of ``\HI\ narrow self--absorption,'' or HINSA \citep{Li03}.  The very close tracking of the \co, \coo, and \CI\  emission velocities with that of the minimum of the \HI\ intensity (see Figure \ref{spectra}), is a strong indication that this is indeed absorption produced by the cold H$^0$ in the central region of L1599B.  

If we assume that the absorption is produced by atomic hydrogen at a temperature of 12 K, we can determine the column density of ``cold'' atomic hydrogen to be 1--3 $\times$10$^{19}$ \cmc.  This value is within the range found in cold dark clouds by \citet{Li03}.  If this cold H$^0$ is located in a path length of 1-2 $\times$10$^{18}$ cm determined from the size of the \coo\ emission, the density of cold H$^0$ must be $\simeq$ 10 \cmv.  This density is significantly higher than the n(\HI) $\simeq$ 1--2 \cmv\ expected in steady--state from  cosmic ray destruction of H$_2$ \citep{Goldsmith05}.  For the density of L1599B, achieving steady state requires approximately 10$^7$ yr, somewhat longer than the age of a few My estimated for the $\Lambda$ Ori ring \citep{Maddalena87}.  It is thus not surprising that the density of cold atomic hydrogen is significantly greater than the steady state value, so that the HINSA interpretation of the \HI\ profiles is plausible.  In this picture, the L159B cloud is a ``transition'' object between purely atomic and fully molecular form.

\section{Discussion}
\label{discussion}

In this section we discuss the kinematics of the L1599B cloud, and analyze the cloud structure focusing on the photochemistry of the cloud edges.  

\subsection{Cloud Kinematics}
\label{kinematics}

\begin{center}
\begin{figure}[h!]
\includegraphics[width = 0.6\textwidth]{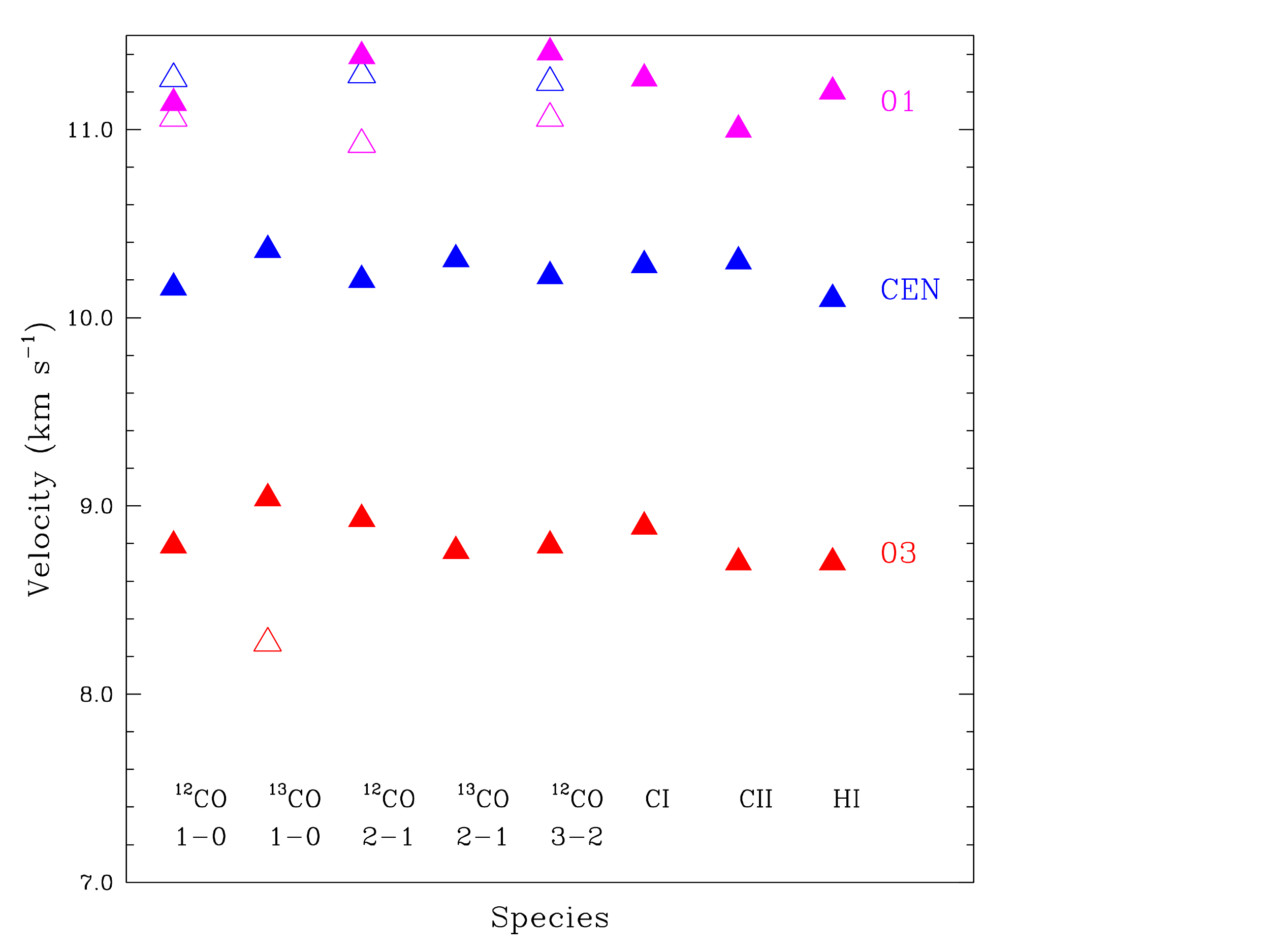}
\caption{\label{velocities} Velocities of peak emission for \co, \coo, \CI, and \CII\ lines observed, and of minimum in the \HI\ line, observed  at the three central positions in L1599B.  The solid symbols are for cases where there is a single component at a given position or the stronger of two components in the cases where a second component is present.  The open symbols denote the velocities of the second, weaker component where present.  The uncertainties in the fitted velocities are given in Table \ref{datatable}, and are typically the size of the plot symbols.  The uncertainty in the \HI\ minimum velocity is more difficult to determine and is a few tenths of a \kms. }
\end{figure}
\end{center}

The spectra in Figure \ref{spectra} and the fitted velocities of the peak emission in Table \ref{datatable} show clear evidence of large--scale kinematic structure in the transverse cut through L1599B.  In Figure \ref{velocities} we plot the peak emission velocities and the velocity of the \HI\ minimum (HINSA interpretation) for the three central positions.  

The key parameters of the different lines at each position follow each other remarkably closely, generally agreeing in velocity within the fitted uncertainty.  It is interesting that the \cplus\ has line parameters similar to those of C$^0$ and CO even though it is almost certainly present only in the ``envelope'' surrounding the atomic and molecular constituents of the cloud.  The lack of velocity shifts between the \cplus\ and the molecular component traced by CO argues against any clear motion of the exterior portion of the cloud relative to its center, as would be expected if the envelope were expanding away from or contracting onto the cloud.   

Such a lack of relative velocity shifts among the species we have observed could be explained if the emission were largely coming from an ensemble of clumps or condensations, with the outer layer of each responsible for the \CII\, and successively more highly shielded layers providing the \CI\ and carbon monoxide.  The overall ensemble of such condensations could exhibit the observed velocity changes. The presence of significant atomic hydrogen in excess of that expected in steady state conditions (discussed in Section \ref{HI} above) argues against the dominance of high density clumps that would have very short timescales for conversion of H$^0$ to H$_2$.  The marginal detection of very weak \hcop\ J = 1 -- 0 emission (\S \ref{HCO+}) suggests that high--density condensations are relatively unimportant, but since we have only this single transition, we cannot make an unambiguous determination of possible clump densities and filling factors.  The observational data on L1599B seem at this point to be satisfactorily reproduced by a smoothly varying density profile and central density much greater than those derived from the carbon monoxide observations discussed in \S\ref{co}.

The \citet{Maddalena87} expanding ring model for the $\Lambda$ Ori region indicates that with an average velocity of 10 \kms\ compared to the  6 \kms\ central velocity of the ring, L1599B is located on the more distant portion of the ring.  The 01 position is closer to $\Lambda$ Ori, and its velocity is red--shifted by 2 \kms\ relative to that of the O3 position.  If we take the CEN position as a reference, the 01 position is red--shifted by 1 \kms\ and the 03 position is blue--shifted by the same amount. If these velocity differences  are due to differential motions within the cloud, it appears to be contracting. While this may appear to be inconsistent with the suggestion of \citet{Andersson93} that the \HI\ halo of L1599B is expanding, their measurements of \HI\ were of the enhanced 21 cm emission surrounding the cloud and not to the absorption features we have observed that presumably trace the denser, primarily molecular cloud core.  

\begin{center}
\begin{figure}[h!]
\includegraphics[width = 0.5\textwidth]{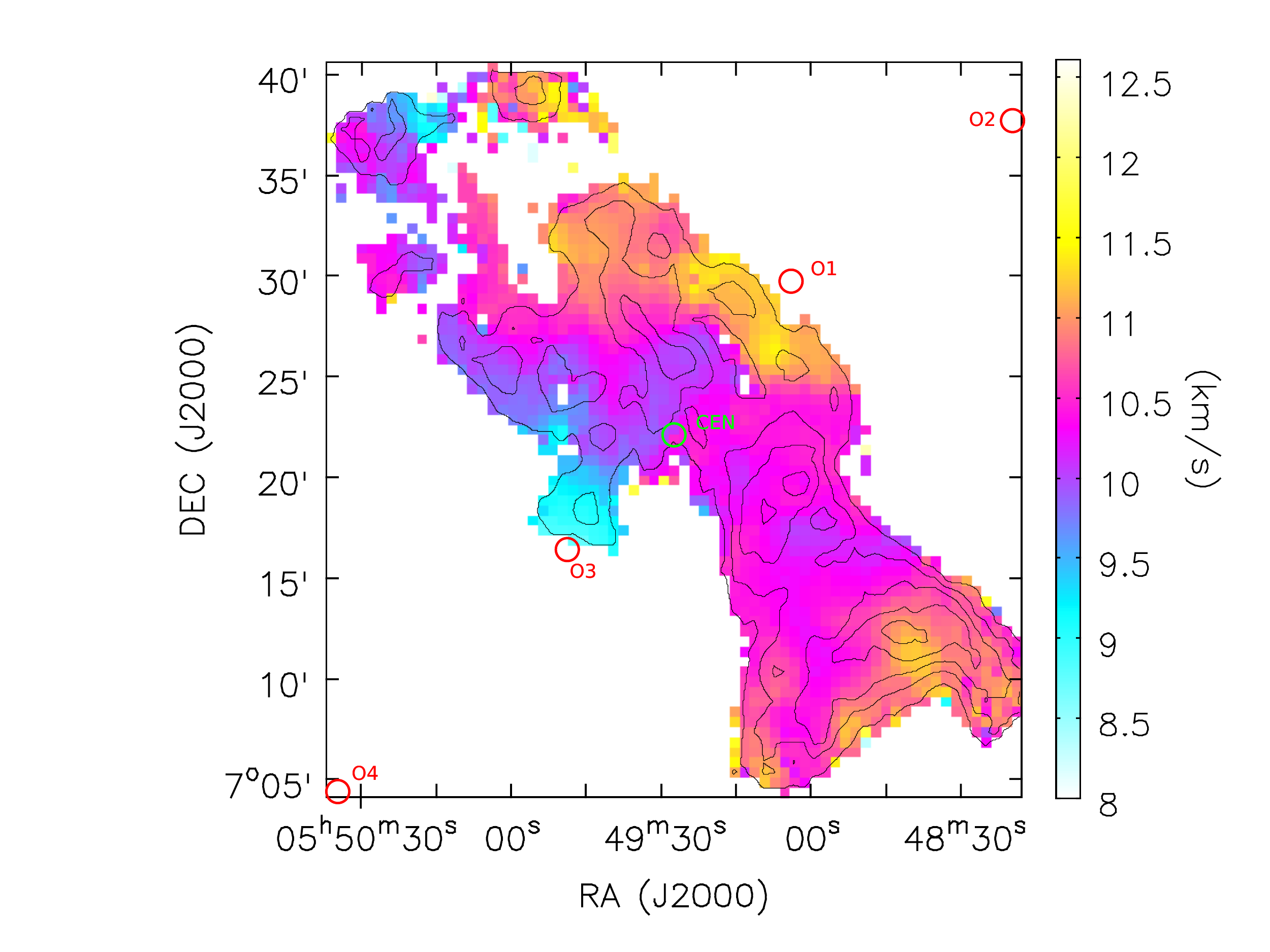}
\caption{\label{13co_vels} Moment 1 image of \coo\ in L1599B with velocities displayed according to the color bar at the right.   The positions studied in detail are indicated.  The main beam temperature cutoff is 3 $\times$ the rms value of 0.23 K.  The black contours show the main beam temperature integrated from 7 \kms\ to 13 \kms.  The contour levels are from 10\% to 90\% in steps of 20\% of the peak value of 7 K \kms. }
\end{figure}
\end{center}

Figure \ref{13co_vels} is a first moment image of \coo\ in L1599B showing the overall behavior of the velocity field.  While there is a velocity shift from one side of the cloud to the other, a significant gradient is largely restricted to the central portion of the cloud, which includes the strip studied in detail.  However, the quite different behavior seen in the southwest portion suggests that while the cloud may have been compressed by the \HII\ region and $\Lambda$ Ori, it is apparent that the overall velocity field is  complex.   Whether the denser portion of the cloud is  contracting while the halo is expanding cannot yet be answered.

The velocity field between $\delta$(J2000) = 7\degr20\arcmin\ and 7\degr30\arcmin\ seen in Figure \ref{13co_vels} shows a velocity gradient which is consistent with rotation of the central portion of the cloud about its long axis.  This region includes the positions studied in detail with the separation of the 01 and 03 positions being 2.15 pc at the assumed distance of 425 pc.  The velocity difference of 2.4 \kms\ corresponds to a velocity gradient of 1.1 \kms\ pc$^{-1}$.  If interpreted as rigid body rotation, the angular velocity $\omega$ = 3.4$\times$10$^{-14}$ s$^{-1}$, which is within the range found for a number of rotating clouds \citep[e.g.][]{Arquilla86}.  However, this behavior is restricted to the central portion of the cloud

\subsection{Cloud Edges}
\label{edges}

\begin{center}
\begin{figure}[h!]
\includegraphics[width = 0.5\textwidth]{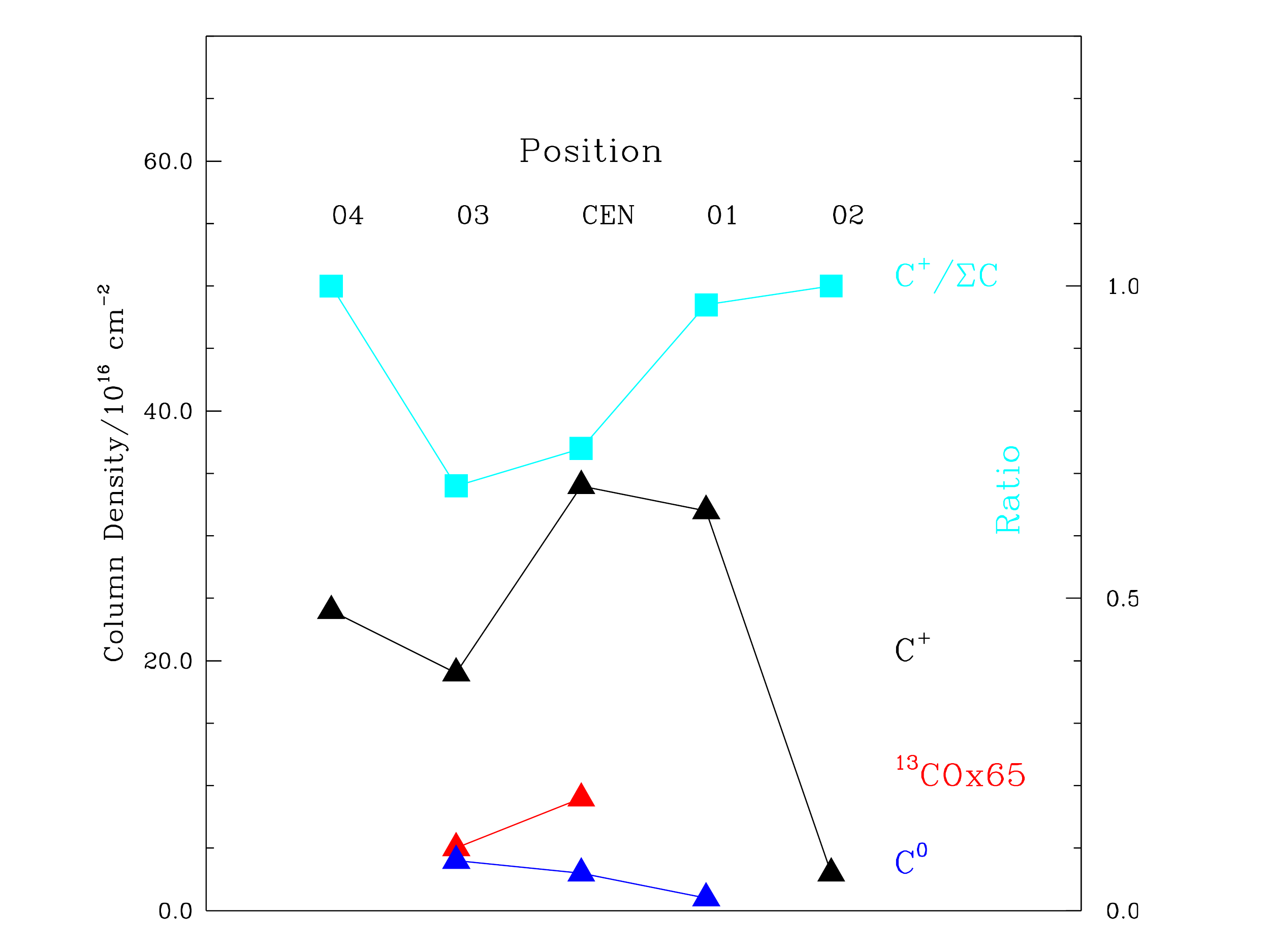}
\caption{\label{coldensfig} Schematic of the distribution of column densities at the 5 positions cutting through the L1599B cloud.   }
\end{figure}
\end{center}

While the observed positions only partially sample the structure of L1599B and are not perfectly aligned with the cloud due to the initial lack of high resolution molecular data, comparing position 02 (towards the star) with 03 (away from the star), we do see an enhancement of the \CII\ emission from the edge of the cloud facing $\Lambda$  Ori.  Without fully--sampled maps in all species we need to be cautious in our conclusions.   We summarize the column density information from \S\ref{coldens} in Table \ref{columns}, in which the column densities have been rounded off to the nearest integer multiple of 10$^{16}$ cm$^{-2}$.  The column densities and the ratio of $N$(C$^+$)/$N$($\Sigma$C), where $N$($\Sigma$C)=$N$(\cplus)+$N$(C$^0$)+65$N$(\coo) is the total carbon column density are shown in Figure \ref{coldensfig}.   Ionized carbon dominates throughout the cloud, but most strongly at the cloud edges.  The enhanced \CII\ emission results from the combination of increased gas temperature and the increased ionized carbon abundance, produced by the stronger radiation field.  The \CII\ line, with $\Delta E/k$ = 91.2 K, is very sensitive to the gas temperature found in cloud edges subjected to radiation fields with modest enhancement factors relative to the standard ISRF.

\begin{table}[t!]
\caption{Column Densities and Ratios in L1599B}
\begin{center}
\label{columns}

\begin{tabular}{cccccc}
\hline\hline & \multicolumn{5}{c}{Position}\\
& 04  & 03  & CEN  & 01  & 02\\
\hline
Species & \multicolumn{5}{c}{Column Density}\\
or Ratio & \multicolumn {5}{c} {(10$^{16}$ cm$^{-2}$)}\\
\hline
\coo\ x 65       & -    & 5    & 9    &  -      & \\
C$^0$            & -    & 4    & 3    & 1      &  - \\
C$^+$            & 24 & 19  & 34  & 32    & 3\\
$\Sigma$C    & 24 & 28  & 46  & 33    & 3\\
C$^+$/C$^0$ &-    & 5   & 11   & 32    & -\\
\hline
\end{tabular}
\end{center}
\end{table}

\begin{figure}[h!]
\includegraphics[width = 0.5\textwidth]{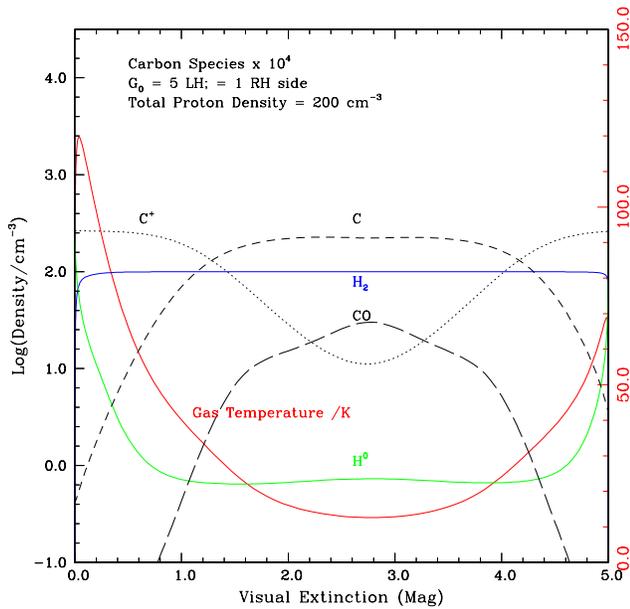}
\caption{\label{pdrfig} Output of {\it PDRLight} model of two--sided slab with the \av\ = 0 mag. side exposed to an interstellar radiation field enhanced by a factor of 5 relative to the standard ISRF, which is incident on the \av\ = 5 mag. side.  The total proton density is 200 \cmv, and the cosmic ray ionization rate is 5$\times$10$^{-17}$ s$^{-1}$.  This is intended to model only the cloud boundaries where the \CII\ emission originates and where the radiation field affects the temperature.  The total extinction is larger than derived from NIR extinction and the central density is far greater than that here, as discussed in the text.}
\end{figure}

To examine this issue in more detail, we have used the {\it PDRLight} version of the Meudon PDR code \citep{Lepetit06}.  We are not modeling the entire cloud, but rather only the cloud boundaries responsible for the \CII\  emission.  We thus adopt a total proton density of 200 \cmv, derived above as reasonable value for this portion of the L1599B cloud.  The model is a plane parallel slab with radiation field enhanced by a factor of 5 on one side relative to the standard interstellar radiation field (ISRF) incident on the other side.  The model computes the steady--state thermal balance and species abundance as a function of optical depth, which is chosen to be 5 magnitudes through the slab.  Due to the modest densities and relative youth of the L1599B cloud, we do not consider any grain surface depletion.  The results are shown in Figure \ref{pdrfig}.  

The most significant results, as shown in Figure \ref{pdrfig}, are that the cloud surface exposed to the enhanced radiation field is warmer, 120 K {\it vs} 70 K at the boundaries, and that the \cplus\ layer is thicker.  The greater thickness results in somewhat larger \cplus\ column density, 2.3 {\it vs} 1.8 $\times$10$^{17}$ \cmc\ within 1 mag. of the cloud surface.  The higher temperature results in much larger fractional population of the upper level of the transition; $f_u$ = 0.028 in the enhanced ISRF side compared to 0.012 in the standard ISRF side.  The combination results in a column density in the $^2$P$_{3/2}$ state equal to 6.4 $\times$10$^{15}$ \cmc\ compared to 2.1 $\times$10$^{15}$ \cmc\ on the low--radiation field side, a factor of 3.1 increase.  The total column densities of \cplus\ necessarily differ slightly relative to those given in \S\ \ref{cplus} that were derived assuming a single set of cloud boundary conditions, but the overall agreement is satisfactory.

The model does not satisfactorily explain all aspects of our data.  In particular, the present model is not realistic in that the total visual extinction is  larger than derived from stellar reddening, computed from infrared reddening using the NICER code \citep{Lombardi01,Chapman07} following the procedure described in \citet{Pineda10}.  This technnique yields a maximum extinction of 1.25 mag. 
Such low extinction regions are ``translucent" clouds having ionized carbon on the outside, and neutral carbon, or a mix of neutral and ionized carbon, on the inside, with little or no CO.
The consequence of the low extinction is a much lower predicted \coo\ column density than observed.  This result may well be connected to the well--known problem of the formation rate of carbon monoxide in diffuse clouds \citep[see  e.g. the discussion in][]{goldsmith13}.  Another issue is that with an extinction as low as indicated by submillimeter and infrared observations, the kinetic temperature (even for n(H$_2$) as high as 1500 \cmv) does not fall below 30 K, significantly higher than that implied by the three observed $^{12}$CO transitions discussed in \S\ \ref{co} above.

The \CI\ emission peaks towards the 03 position, in the direction away from the star.  Using the column densities  from \S\ \ref{CI} and \ref{cplus}, the ratio $N({\rm C}^+)/N({\rm CO})$ is 32, 12, and 5, at positions 01, CEN, and 03, moving from the side of L1599B facing $\Lambda$ Ori to the side facing away.  This change is largely due to the reduction of the column density of neutral carbon by the enhanced radiation field produced by the star, together with a modest enhancement of the ionized carbon column density.  

\section{Summary}
\label{summary}

We have carried out an observational study of the L1599B cloud, located in the ring surrounding the O III star $\Lambda$ Ori.  We have made images in the J = 1--0 transitions of \co\ and \coo\, but focus on five positions for which we have an extensive set of spectral diagnostics of the gas properties, cutting through the cloud on a line directed towards the star.  At these positions we have observed the 2--1 transitions of \co\ and \coo, the 3--2 transition of \co, the $^3$P$_1$--$^3$P$_0$ transition of \CI, the $^2$P$_{3/2}$--$^2$P$_{1/2}$ transition of \CII, and used 21 cm \HI\ profiles from the GALFA survey.  All are seen in emission with the possible exception of the \HI, which may be showing self--absorption by cold atomic gas in the molecular cloud.  

The line centroid velocities and also the line widths for the emission lines agree very closely, and all exhibit a systematic velocity shift as one moves across the cloud.  This shift could be indicative of rotation but is more likely the result of contraction of the cloud, comprising molecular, atomic, and ionic components, due to compression by the \HII\ region

Using the two lowest transitions of \coo, we have determined the H$_2$ volume density in the shielded portion of the cloud to be 1000 and 2500 \cmv\ at the 03 and CEN positions, respectively.  This range delineates the density in the central portion of the cloud, where from the three lowest transitions of \co, we find kinetic temperatures of 12 and 11 K, respectively.
  
The visual extinction through the cloud is modest, $\simeq$ 1.25 mag, and the \coo\ column densities are 0.7-1.3$\times$10$^{15}$ \cmc.  These column densities are greater than expected from a PDR model of the L1599B cloud which includes a radiation field enhancement of a factor of 5 on the side facing the star $\Lambda$ Ori.  The \coo\ emission peaks on the side of L1599B facing away from L1599B.

The \CII\ emission is enhanced and the \CI\ emission reduced at the position of the cloud facing $\Lambda$ Ori, and the ratio $N({\rm C}^+)/N({\rm CO})$ at positions 01, CEN, and 03, moving from the side of L1599B facing $\Lambda$ Ori to the side facing away is 32, 12, and 5.  While comparisons with a uniform density slab PDR model should be treated with caution, it appears that the model does predict the enhancement of \cplus\ and the diminution of C$^0$ reasonably well.  While the \cplus\ emission from regions with such modest radiation field is obviously weak, our study confirms that relatively extended regions even quite far from massive young stars can contribute significantly to the \CII\ emission and should be included in calculating the total \cplus\ luminosity and cooling of star--forming regions.

\begin{acknowledgements}

We  express our gratitude to the Staff at Qinghai Station of the Purple Mountain Observatory for carrying out the observations of the 1--0 transitions of carbon monoxide.  We are grateful to the staff of the Korean VLBI Network (KVN).   The KVN is a facility operated by the Korea Astronomy and Space Science Institute.  We thank Josh Peek and Marko Krco for assistance with the \HI\ data and its interpretation, and B.--G.  Anderson for discussions about the \HI\ in L1599B.  We are grateful to Franck Le Petit and Jacques Le Bourlot for assistance with using the latest version of the Meudon PDR code.  An anonymous referee made very helpful comments that improved the paper.
This research was conducted at the Jet Propulsion Laboratory, which is operated by the California 
Institute of Technology under contract with the National Aeronautics and Space Administration (NASA). 
\copyright2016 California Institute of Technology. \\

\end{acknowledgements}

\bibliography{bibdata}
\end{document}